\documentclass[twocolumn,tighten]{aastex63}

\usepackage{graphicx}
\usepackage{amsmath}
\usepackage{amssymb}
\usepackage{color}
\usepackage{mathtools}
\usepackage{ulem}
\usepackage[all]{hypcap} 

\usepackage{lineno}

\newcommand\msun{\, {M}_\odot}

\def\apjl{ApJL}%
\def\apjs{ApJS}%
\def\aap{A\&A}%
\begin{document}

\title{Neutron star kicks and implications for their rotation at birth}
\shorttitle{Neutron star kicks and rotation at birth}
\correspondingauthor{Giacomo Fragione}
\email{giacomo.fragione@northwestern.edu}

\author[0000-0002-7330-027X]{Giacomo Fragione}
\affil{Center for Interdisciplinary Exploration \& Research in Astrophysics (CIERA), Northwestern University, Evanston, IL 60208, USA}
\affil{Department of Physics \& Astronomy, Northwestern University, Evanston, IL 60208, USA}

\author[0000-0003-4330-287X]{Abraham Loeb}
\affil{Astronomy Department, Harvard University, 60 Garden St., Cambridge, MA 02138, USA}

\begin{abstract}
Neutron stars are born out of core-collapse supernovae, and they are imparted natal kicks at birth as a consequence of asymmetric ejection of matter and possibly neutrinos. Unless the force resulting from the kicks is exerted exactly at their center, it will also cause the neutron star to rotate. In this paper, we discuss the possibility that neutron stars may receive off-center natal kicks at birth, which imprint a natal rotation. In this scenario, the observed pulsar spin and transverse velocity in the Galaxy are expected to correlate. We develop a model of the natal rotation imparted to neutron stars and constrain it by the observed population of pulsars in our Galaxy. When considering a single-kick position parameter, we find that the location of the off-center kick is $R_{\rm kick}=2.03^{+3.74}_{-1.69}$\,km at $90\%$ confidence, and is robust when considering pulsars with different observed periods, transverse velocities, and ages. Nonetheless, the model encounters challenges in effectively fitting the data, particularly at small transverse velocities, prompting the exploration of alternative models that include more complex physics. Our framework could be used as a guide for core-collapse simulations of massive stars.
\end{abstract}

\section{Introduction}
\label{sect:intro}

Since the discovery of the first pulsar \citep{HewishBell1968}, the number of known neutron stars has grown to more than three thousands, including a rich ensemble of binary and millisecond pulsars \citep{ManchesterHobbs2005}\footnote{\url{https://www.atnf.csiro.au/research/pulsar/psrcat/}}. The study of pulsars combines a wide range of physical processes, which shed light on the fundamental laws of nature and the structure of our Galaxy \citep[e.g.,][]{LorimerKramer2004,LorimerPol2019}. Timing of pulsars in binary systems has not only allowed exquisite tests of strong gravity, but has also given insights into the equation of state that describes them and the nature of the binary systems they inhabit \citep[e.g.,][]{HulseTaylor1975,ThompsonDuncan1993,ThorsettChakrabarty1999,LyneBurgay2004,KramerStairs2006,AntoniadisFreire2013,CromartieFonseca2020}.

Neutron stars form as a result of the core collapse of stars with initial masses approximately in the range $[8-20]\msun$ \citep[for a review see][]{WoosleyHeger2002}. The associated supernova explosions eject asymmetrically matter (and possibly neutrinos) at velocities of up to $\sim 10^4$\,km\,s$^{-1}$ \citep[e.g.,][for a review]{Janka2012,Muller2020}. Therefore, conservation of momentum implies that neutron stars are imparted a significant natal kicks, as observed for pulsars in our Milky Way. The natal kick velocities imparted to neutron stars play a key role in a number of physical processes, from their retention in globular clusters to the evolution of binaries, including X-ray binaries and gravitational wave sources \citep[e.g.,][]{BhattacharyavandenHeuvel1991,BrandtPodsiadlowski1995,JonkerNelemans2004,FryerBelczynski2012,TaurisKramer2017,YeFong2020,FragioneLoeb2021}.

Over the last twenty years, a number of studies have inferred the distribution of natal kicks imparted to neutron stars at birth from the increasingly large and accurately measured sample of observed single pulsar velocities \citep[e.g.,][]{HarrisonTademaru1975,HansenPhinney1997,HobbsLorimer2005,ArzoumanianChernoff2002,KapilMandel2023}. Despite the various mean velocities predicted by different studies, a common conclusion has been that neutron stars are typically imparted natal kicks of the order of several hundreds km\,s$^{-1}$. 

The specific physical process that kicks the neutron star at birth is still highly debated \citep[e.g.,][]{FryerKusenko2006,OttBurrows2006,IgoshevPopov2013,MaFuller2019,NagakuraSumiyoshi2019,ColemanBurrows2022,JankaWongwathanarat2022}. Unless its force is exerted exactly at their center, any kick mechanism will also cause the neutron star to rotate, as originally suggested by \cite{SpruitPhinney1998} on purely theoretical grounds.

The idea that natal kicks could imprint a spin at birth has been explored only in a few cases. For example, \citet{FarrKremer2011} considered this scenario to explain the spin tilts in the double pulsar PSR J0737-3039. However, a study of the role of off-center kicks in neutron stars at birth has never been performed at a population level. In this paper, we consider for the first time the possibility that natal kicks can impart a natal rotation to neutron stars, and constrain it using current observations on Galactic pulsars.

This paper is organized as follows. In Section~\ref{sect:method}, we discuss how natal kicks can impart a natal rotation to neutron stars. In Section~\ref{sect:results}, we present our results and show that neutron stars receive kicks at about $1$\,km from their center at a population level. Finally, in Section~\ref{sect:concl}, we discuss the implications of our results and draw our conclusions.

\section{Off-center kicks}
\label{sect:method}

First, we describe how off-center natal kicks can impart a spin to neutron stars at birth.

The spin at birth can be modeled as the sum of two contributions. The first contribution is the spin that the neutron star may inherit from the rotation of the progenitor star before explosion
\begin{equation}
    \mathbf{S_{\rm b,*}} = \frac{2\pi}{P_{\rm b,*}} \mathbf{\hat{j}}\,,
\end{equation}
where $\mathbf{\hat{j}}$ is the rotation axis and $P_{\rm b,*}$ its period. The second contribution arises from the possible off-center natal kick imparted to the neutron star at formation, as a result of asymmetries associated with the supernova ejecta
\begin{equation}
    \Delta \mathbf{S} = \frac{M_{\rm NS}}{I_{\rm NS}} \mathbf{R}_{\rm kick} \times \mathbf{V}_{\rm kick}\,.
\end{equation}
Here, $M_{\rm NS}$ and $I_{\rm NS}$ are the pulsar mass and moment of inertia, which we fix to $1.4\msun$ and $50\msun\ {\rm km}^2$ \citep[e.g.,][]{WoosleyHeger2002}, respectively, $V_{\rm kick}$ is the velocity kick, and $R_{\rm kick}$ is the location of the kick with respect to the center. Therefore, the total spin of a neutron star at birth is simply
\begin{equation}
    \mathbf{S_0}=\mathbf{S_{\rm b,*}} + \Delta \mathbf{S}\,.
\end{equation}
Regarding its magnitude, it is straightforward to show that
\begin{equation}
    S_0 = \left[\left(S_{\rm b,*}+\Delta S \cos\gamma \right)^2+\left(\Delta S \sin\gamma \right)^2\right]^{1/2}\,,
    \label{eqn:s0}
\end{equation}
where $\gamma$ is the angle between $\mathbf{S_{\rm b,*}}$ and $\Delta \mathbf{S}$, and
\begin{equation}
\Delta S = \frac{M_{\rm NS} R_{\rm kick} V_{\rm TR} \sin\alpha}{I_{\rm NS} \sin\beta}\,.
\end{equation}
Here, $\alpha$ is the angle between $\mathbf{R}_{\rm kick}$ and $\mathbf{V}_{\rm kick}$, and $V_{\rm TR} = V_{\rm kick} \sin\beta$ to correct for the fact the we only measure the 2D transverse velocity of pulsars (in Galactic coordinates), $V_{\rm TR}$, with $\beta$ being the projection angle. Hence, if pulsars receive an off-center kick, there should exist a correlation between their period at birth and the magnitude of the natal kick velocity.

Note that we observe pulsars not right after they are born, rather at some time after their birth. Therefore, we need to account for the fact that pulsars spin down during their lifetime as a result of energy emission through dipole magnetic braking \citep{GoldreichJulian1969,Sturrock1971}. The spin of a neutron star at birth, $P_0=2\pi/S_0$, is related to the its spin, $P$, at some time $\tau$ after birth through
\begin{equation}
    \frac{P^2-P_0^2}{2} = P\dot{P} \tau\,,
    \label{eqn:pp0}
\end{equation}
where $\dot{P}$ is the rate of spin down. From observations we can only directly estimate the characteristic age
\begin{equation}
    \tau_{\rm ch} = \frac{P}{2\dot{P}}\,
\end{equation}
of a pulsar, which represents an upper limit to the pulsar true age. Introducing $\tilde{\tau} = \tau/\tau_{\rm ch}$, which is naturally defined in the range $[0-1]$, Equation~\ref{eqn:pp0} can be recast as
\begin{equation}
    P=\frac{2\pi}{S_0(1-\tilde{\tau})^{1/2}}\,.
    \label{eqn:fit}
\end{equation}
The above equation relates the observed pulsars spin and transverse velocity, and can be used to constrain the parameters of our model. 

\section{Statistical constraints from pulsar catalogs}
\label{sect:results}

We use a Bayesian framework to determine the possible off-center distance of kicks imparted at birth to neutron stars. The likelihood of observing a given period of a pulsar, $P_i$, given a model (Equation~\ref{eqn:fit}) is
\begin{equation}
    p({\pmb \theta}|{\pmb P}, {\pmb V_{\rm TR}}) \propto \pi({\pmb \theta}) \prod_i p(P_i|{\pmb \theta}, V_{\rm TR,i})\,,
\end{equation}
where $\pmb \theta=\left\{R_{\rm kick}, \alpha, \beta, \gamma, P_{\rm birth}, \tilde{\tau}\right\}$ is the set of parameters describing our model and $\pi({\pmb \theta})$ is the priors on our parameters. We use uniform priors for $R_{\rm kick}$ in the range $[0-10]$\,km, $P_{\rm birth}$ in the range $[0-1]$\,s, and $\tilde{\tau}$ in the range $[0-1]$, while we use isotropic priors for the angles $\alpha$, $\beta$, $\gamma$. We use the nested-sampling code \textsc{nestle}\footnote{\url{http://kylebarbary.com/nestle/index.html}} to maximize the log-likelihood of our model and to infer the confidence regions of our parameters. 

\begin{table}
\caption{Minimum period ($P_{\min}$), minimum transverse velocity ($V_{\rm TR, min}$), and maximum age ($\tau_{\rm ch, max}$) of our cuts in our analysis of the pulsar catalog data. The radius ($R_{\rm kick}$) provides the median and $90\%$ confidence regions for the location of the off-center kick.}
\centering
\begin{tabular}{ccccc}
\hline\hline
$P_{\min}$ (s) & $V_{\rm TR, min}$ (km s$^{-1}$) & $\tau_{\rm ch, max}$ (Myr) & $R_{\rm kick}$ (km)\\
\hline
$0.01$ & $100$ & $10^4$ & $2.03^{+3.74}_{-1.69}$\\
$0.01$ & $150$ & $10^4$ & $1.89^{+4.27}_{-1.67}$\\
$0.01$ & $200$ & $10^4$ & $2.03^{+4.20}_{-1.84}$\\
$0.01$ & $250$ & $10^4$ & $1.77^{+4.18}_{-1.66}$\\
$0.01$ & $100$ & $10^3$ & $2.32^{+4.41}_{-1.92}$\\
$0.01$ & $150$ & $10^3$ & $1.89^{+3.74}_{-1.65}$\\
$0.01$ & $200$ & $10^3$ & $2.02^{+4.20}_{-1.84}$\\
$0.01$ & $250$ & $10^3$ & $1.77^{+4.18}_{-1.66}$\\
$0.01$ & $100$ & $10^2$ & $1.66^{+3.61}_{-1.45}$\\
$0.01$ & $150$ & $10^2$ & $1.98^{+4.26}_{-1.73}$\\
$0.01$ & $200$ & $10^2$ & $1.96^{+4.09}_{-1.79}$\\
$0.01$ & $250$ & $10^2$ & $1.65^{+3.98}_{-1.55}$\\
$0.01$ & $100$ & $10^1$ & $1.94^{+4.03}_{-1.70}$\\
$0.01$ & $150$ & $10^1$ & $2.10^{+4.41}_{-1.84}$\\
$0.01$ & $200$ & $10^1$ & $2.46^{+4.69}_{-2.20}$\\
$0.01$ & $250$ & $10^1$ & $2.42^{+4.91}_{-2.19}$\\
$0.1$  & $100$ & $10^4$ & $1.91^{+3.93}_{-1.59}$\\
$0.1$  & $150$ & $10^4$ & $1.75^{+3.89}_{-1.55}$\\
$0.1$  & $200$ & $10^4$ & $1.90^{+4.40}_{-1.74}$\\
$0.1$  & $250$ & $10^4$ & $1.58^{+4.02}_{-1.49}$\\
$0.1$  & $100$ & $10^3$ & $2.06^{+3.86}_{-1.70}$\\
$0.1$  & $100$ & $10^3$ & $1.75^{+3.89}_{-1.55}$\\
$0.1$  & $100$ & $10^3$ & $1.90^{+4.40}_{-1.74}$\\
$0.1$  & $100$ & $10^3$ & $1.58^{+4.02}_{-1.49}$\\
$0.1$  & $100$ & $10^2$ & $1.57^{+3.58}_{-1.38}$\\
$0.1$  & $150$ & $10^2$ & $1.66^{+3.49}_{-1.45}$\\
$0.1$  & $200$ & $10^2$ & $1.73^{+4.15}_{-1.60}$\\
$0.1$  & $250$ & $10^2$ & $1.49^{+3.73}_{-1.41}$\\
$0.1$  & $100$ & $10^1$ & $1.74^{+3.38}_{-1.52}$\\
$0.1$  & $150$ & $10^1$ & $1.87^{+4.09}_{-1.68}$\\
$0.1$  & $200$ & $10^1$ & $2.21^{+4.63}_{-1.99}$\\
$0.1$  & $250$ & $10^1$ & $2.22^{+4.82}_{-2.03}$\\
\hline\hline
\end{tabular}
\label{tab:models}
\end{table}

For pulsar data, we use the Australia Telescope National Facility Pulsar Catalogue \citep{ManchesterHobbs2005}, which contains information on the observed period of pulsars as a function of their observed transverse velocity. In order to have a cleaner sample, we remove from our catalog anomalous X-ray pulsars. We also discard pulsars in binaries, where the details of the specific binary evolution path could affect the measured pulsar periods and transverse velocity. Finally, we also remove pulsars associated with globular clusters from our sample since dynamical interactions would have naturally affected their observed transverse velocity. 

\begin{figure} 
\centering
\includegraphics[scale=0.575]{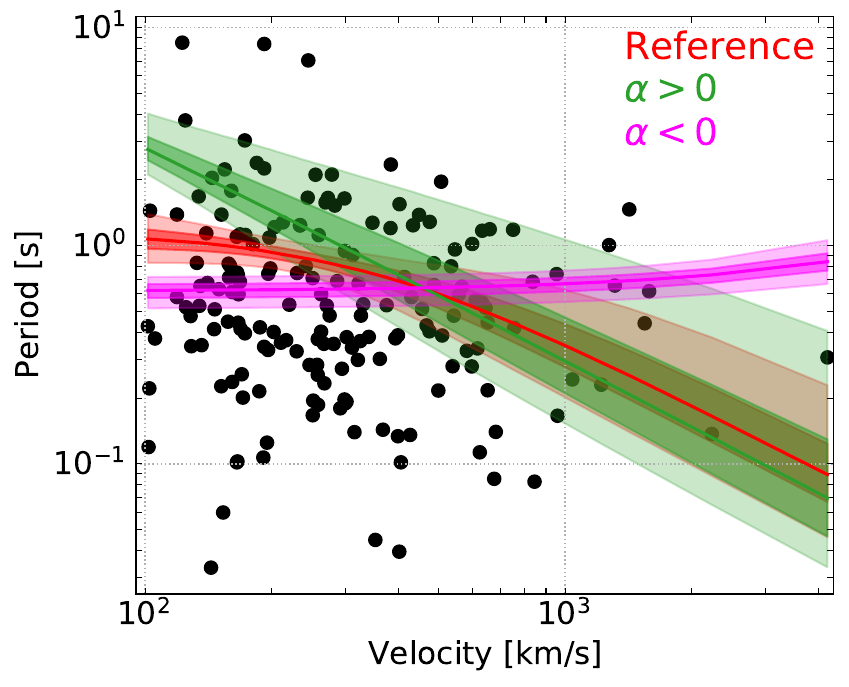}
\caption{Observed period of pulsars as a function of their observed transverse velocity ($P_{\min}=0.01$\,s, $V_{\rm TR, min}=100$\,km\,s$^{-1}$, and $\tau_{\rm ch, max}=10^4$\,Myr). The red solid line shows the best-fit model in our analysis. The green and magenta lines show the best fit in the case the location of the kick is described by a positive or negative power-law distribution of the distance from the center of the neutron star, respectively. The dark red and light bands denote the $50\%$ and $90\%$ credibility regions.}
\label{fig:fit}
\end{figure}

Figure\,~\ref{fig:fit} shows the observed period of pulsars as a function of their observed transverse velocity for our sample. We report the observed period of pulsars as a function of their observed transverse velocity, assuming a minimum pulsar period of $P_{\min}=0.01$\,s, and a minimum transverse velocity of $V_{\rm TR, min}=100$\,km\,s$^{-1}$, without any restrictions on the pulsars age.

If there is an off-center kick, the observed period and transverse velocity of a pulsar should be inversely proportional to each other. In Figure\,~\ref{fig:fit}, we also report the results of our statistical analysis. We show the the best-fit model in our analysis (red solid line) and the $50\%$ and $90\%$ confidence regions (dark red and light red bands, respectively). Our results show that there is indeed some correlation between the pulsars observed periods and transverse velocities.

\begin{figure*} 
\centering
\includegraphics[scale=0.525]{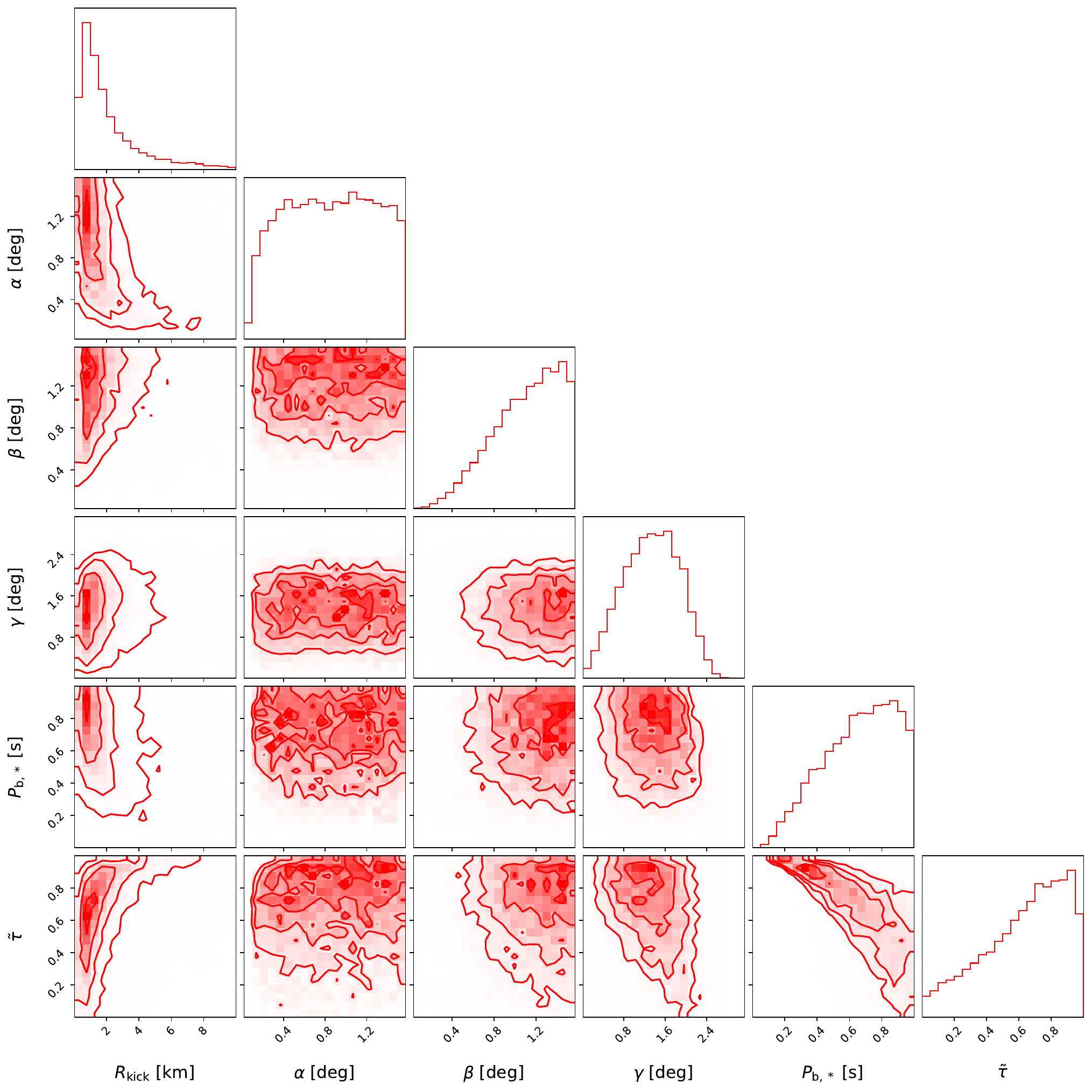}
\caption{Hyper-posterior for parameters describing the relation between the observed period of pulsars and their observed transverse velocity (Equation~\ref{eqn:fit}).}
\label{fig:corner}
\end{figure*}

This is clear also from Figure~\ref{fig:corner}, where we shows the hyper-posterior for parameters our model. In particular, we find that the location of the off-center kick has a median of about $1$\,km from the neutron star center. At $90\%$ confidence, we find that $R_{\rm kick}=2.03^{+3.74}_{-1.69}$\,km, therefore implying a kick close to the center of the neutron star. We also show that the spin imparted at birth by natal kicks is typically perpendicular ($\gamma$ peaks at about $90^\circ$) to any spin that the neutron star inherits from the rotation of the progenitor star. Finally, we find that pulsars have a wide distribution of ages, with some having an age much smaller than the maximal characteristic age, but with the majority having a true age of about $80\%$ of $\tau_{\rm ch}$.

We repeat our analysis considering different cuts on the observed pulsar periods and transverse velocities to check that our analysis is robust (see Table~\ref{tab:models}). Indeed, the velocity of pulsars could be influenced by the Galactic potential over time. Younger pulsars experience less impact on their velocity, while those with higher velocities are expected to be less affected by the Galactic potential. For the period, the reason of exploring different cuts is that we do not know exactly the boundary between recycled (millisecond) pulsar and non-recycled pulsars \citep[e.g.,][]{TaurisLanger2012}. For the transverse speed, the motivation is that the progenitors of neutron stars are typically found in binaries \citep[e.g.,][]{SanadeMink2012,MoeDiStefano2017}. If pulsars were in binaries at birth, the binary orbital motion could affect the observed transverse velocity. Finally, we also take into account different cuts on the pulsar age, to check if our results are consistent when considering only younger pulsars, before they significantly spin down \citep{GoldreichJulian1969,Sturrock1971}. We find that the median of the location of the off-center pulsar kick is always about $1$\,km from the pulsar center, with also the $90\%$ confidence regions also being consistent. Note that we find a marginal trend with the age of the pulsar: the younger the pulsars are, the further is the location of the kick in our analysis.

We repeat our analysis in the case the location of the kick depends as a power-law distribution on the inner radius of the star
\begin{equation}
    f(R_{\rm kick})\propto R^{\alpha}_{\rm kick}\,,
\end{equation}
in the range $[0-10]$\,km. We plot the best-fit model and the $50\%$ and $90\%$ confidence regions in Figure~\ref{fig:fit} in the cases we fit for $\alpha>0$ (green lines) and $\alpha<0$ (magenta lines). However, we note that for $\alpha<0$ an anti-correlation of the pulsar period and velocity is introduced, which makes this model not particularly meaningful. The best-fit value of the index of the power law is $1.25$ and $-1.45$ in case we assume a positive and negative prior on $\alpha$, respectively. We find that the Bayes ratio between the three different models is close to unity, therefore there is no strong evidence that any of the models is statistically preferred over the others. As shown in Figure~\ref{fig:fit}, our models tend not to reproduce well the observed data, especially at low values of the transverse velocity, probably pointing to the need of including more complex physics in our framework.

\section{Discussion and Conclusions}
\label{sect:concl}

In this paper, we have considered the possibility that neutron stars receive off-center kicks at birth, as a result of asymmetric ejection of matter and neutrinos during supernovae, as originally suggested by \cite{SpruitPhinney1998} on purely theoretical grounds. We have developed a model that describes the relation between the observed spin and transverse velocity of neutron stars, assuming that their rotation at birth is the sum of the rotation inherited by the progenitor star and the rotation imparted by natal kicks. We have constrained the parameters of our model using the Australia Telescope National Facility Pulsar Catalogue \citep{ManchesterHobbs2005}. We have found that the location of the off-center kick is $R_{\rm kick}=2.03^{+3.74}_{-1.69}$\,km from the neutron star center, at $90\%$ confidence. However, we note that our model encounters challenges in fitting the data, especially at low values of the transverse velocity, which encourages the exploration of alternative models, which we leave to future works.

We have also explored how our results would be affected by different cuts on the observed periods, transverse velocities, and ages, as younger and faster pulsars are expected to be less affected by the Galactic potential over time. We find that our results do not depend significantly on the different cuts on the pulsar periods, transverse velocities, and ages. At $90\%$ confidence, our results are still consistent with kicks originating close to the center of the newborn neutron star. Our model and findings are independent from the underlying equation of state, which can however affect the geometry and the physics that originate the kick.

Several studies have utilized X-ray observations of compact nebulae to explore the potential relationship between the proper motion of young pulsars and their spin direction, represented by the symmetry axis of the nebula \citep{LaiChernoff2001,JohnstonHobbs2005,WangLai2006,KuranovPopov2009,NgRomani2004,NgRomani2007,Rankin2007}. According to our analysis, the spin imparted during the pulsar birth from natal kicks tends to be perpendicular to the rotational spin inherited from the progenitor star. As a result, in certain cases, the kick velocity and the rotation spin align in the same direction, consistently with previous findings. However, note that some recent work has questioned the evidence for a possible spin-kick alignment, as predicted by  the mechanism of jittering-jets explosion with accretion of stochastic angular momentum \citep{BearSoker2018,Soker2022}.

As discussed, our model faces challenges in fitting the data, which may point to the need of a more complex framework. For example, our model lacks consideration for complex explosion physics, such as the possibility of multiple kicks, neutron star radius evolution post-kicks, and a neutron star mass spectrum. Additionally, our data source, the ATNF catalog, is a compilation of various search programs with potentially different selection biases, unlike some other catalogs \citep[e.g.,][]{JankowskiBailes2019}. Moreover, the absence of proper motion measurements for many discovered pulsars could impact our analysis. Faster and younger pulsars tend to be observed more frequently and are more likely to have proper motion data, but if the proper motion is not statistically significant, it may not be reported or included in the catalog. Furthermore, the way proper motion is measured, with timing precision proportional to $1/P$, can result in slower velocities being measured for faster pulsars, potentially affecting our results and deviating from the claimed trend.

\section*{Acknowledgements}

We thank David Kaplan, Fred Rasio, Adam Burrows, and Claire S. Ye for insightful discussions. This work was supported by NASA Grant 80NSSC21K1722 at Northwestern University. AL was supported in part by the Black Hole Initiative at Harvard University, which is funded by grants from GBMF and JTF.

\bibliography{refs}

\end{document}